**Spin-related transport in a polycrystalline $NiCo_2O_4$ film: Drastic current-induced change in resistivity–temperature characteristics via spin injection**

Shiho Sugiyama,[1] Masaaki Tanaka,[1,2,3] and Ryosho Nakane[1,3,4]
[1)]*Department of Electrical Engineering and Information Systems, The University of Tokyo, 7-3-1 Hongo, Bunkyo-ku, Tokyo 113-8656, Japan*
[2)]*Center for Spintronics Research Network (CSRN), The University of Tokyo, 7-3-1 Hongo, Bunkyo-ku, Tokyo 113-8656, Japan*
[3)]*Institute for Nano Quantum Information Electronics, The University of Tokyo, 4-6-1 Komaba, Meguro-ku, Tokyo 153-8505, Japan*
[4)]*Systems Design Lab (d.lab), The University of Tokyo, 7-3-1 Hongo, Bunkyo-ku, Tokyo 113-8656, Japan*

We have studied spin-related transport in a polycrystalline $NiCo_2O_4$ (NCO) film on a $MgAl_2O_4$/Si(001) substrate, motivated by potential applications of the theoretical half-metallicity of NCO to Si-based high-performance spin-transport devices. Our approach is to systematically measure and analyze the temperature dependence of the film's resistivity ($\rho-T$) with various in-plane currents (100 nA–1 mA) and temperatures (4–290 K). With increasing current, the $\rho-T$ curve changes drastically from semiconducting ($d\rho/dT<0$) to non-monotonic and eventually toward metallic ($d\rho/dT>0$). A distinctive feature is that the single NCO film exhibits a $\rho-T$ characteristic of polycrystalline defective NCO at 100 nA, whereas it exhibits a $\rho-T$ characteristic of epitaxial less-defective NCO over a wide temperature range at 1 mA. This current-induced evolution of $\rho-T$ reflects the enhancement of the Curie temperature of defective regions near grain boundaries, accompanied by enhanced spin alignment there. We proposed a spin-related transport model that extends conventional hopping conduction models, by incorporating the temperature- and current-dependent degree of spin alignment, as well as its spatial dependence inherent to polycrystalline NCO. This model comprehensively explains the interplay between the spin-alignment profile and transport mechanism. The analysis reveals that spin injection from grain bodies to grain boundaries enhances the spin alignment there and strengthens double-exchange interactions, facilitating conduction. This phenomenon strongly depends on both temperature and current. Our findings provide evidence of spin-polarized electrons inside the grain bodies, highlighting the potential of our polycrystalline NCO film as an efficient spin source. The present model is further supported by current−voltage and magnetoresistance features.

Si-based spin-transport devices are attractive for low-power next-generation electronic devices, because Si has a long spin lifetime[1] and is highly compatible with conventional complementary metal-oxide-semiconductor (CMOS) technology. To realize such devices with high performance, efficient spin injection into Si is essential[2]. A typical approach is to drive a spin-polarized current through a ferromagnetic (FM) metal electrode/insulator/Si substrate[3–8], where the insulator tunnel barrier is inserted to overcome the conductivity mismatch[9] between the FM metal and Si and/or to suppress both the oxidation of the Si surface and the diffusion of magnetic elements into the Si substrate. Such FM electrodes require both sufficient conductivity and high spin polarization, making the conductive, half-metallic, and ferrimagnetic oxide $NiCo_2O_4$ (NCO) a useful candidate. At least 73% spin polarization of NCO was experimentally demonstrated in a fully epitaxial NCO/$MgAl_2O_4$(MAO)/NCO[10]. Therefore, an NCO/MAO/Si is promising for spin injection into Si because it leverages the well-established combination of NCO and MAO. NCO in this structure, however, is expected to be polycrystalline owing to the lattice mismatch between Si and MAO (5.2%[11,12]) or NCO (5.6%[12,13]). Thus, it is necessary to clarify the material properties of polycrystalline NCO films via investigation of their electron transport characteristics that sensitively reflect their microscopic details.

The conductive property and half-metallicity of NCO originate from the transfer of down-spin electrons between Ni cations coupled ferromagnetically by oxygen-mediated double-exchange interactions[14,15]. High-quality epitaxial NCO films on MAO(001) substrates have strong double-exchange interactions, resulting in saturation magnetization $M_s$ close to the theoretical value[16] of $2\mu_B$ / formula unit (f.u.)[17–19] and metallic temperature dependence of resistivity ($\rho-T$) ($d\rho/dT > 0$) up to room temperature[17,20,21], except for the lowest-temperature range[20]. In contrast, NCO films with numerous defects, such as anti-phase boundaries[15] and oxygen vacancies[20], as well as polycrystalline NCO, which generally has many defects particularly near grain boundaries, have weaker double-exchange interactions, thereby exhibiting reduced $M_s$ and semiconducting $\rho-T$ ($d\rho/dT < 0$)[15,20]. Such electron transport in polycrystalline NCO has been typically analyzed using a variable range hopping (VRH)[22] model in a low-temperature range and a nearest neighbor hopping (NNH) model[22] in a high-temperature range to estimate the characteristic temperatures $T_{VRH}$ and $T_{NNH}$[23,24]. These conventional hopping models alone are not sufficient to accurately capture the properties of polycrystalline NCO, since they represent the properties of the entire

NCO sample by a single characteristic temperature and thereby implicitly neglect the spatially inhomogeneous defect distribution intrinsic to polycrystalline NCO. Thus, a more comprehensive transport model incorporating such non-uniformity is necessary.

$La_{1-x}Sr_xMnO_3$ (LSMO) shares with NCO a double-exchange-mediated transport mechanism[25], where a higher degree of spin alignment facilitates electron transfer and decreases $\rho$, and the electron transport in LSMO has been studied more extensively[26–28] than in NCO. Of particular relevance here are studies on spin-related transport in samples with inhomogeneous defect distributions[27,28], where defect-rich regions have a lower degree of spin alignment and a lower Curie temperature $T_C'$ than other nearly defect-free regions. In these studies, non-monotonic $\rho-T$ has a peak near $T_C'$, exhibiting metallic $\rho-T$ below $T_C'$ and semiconducting $\rho-T$ above $T_C'$. Ref. 28 also reported current-induced $T_C'$ enhancement, observable as a shift of the $\rho-T$ peak toward higher temperatures and as an extension of the temperature range with metallic $\rho-T$[28]. In this phenomenon, spin-transfer torque[29] caused by spin-current injection from the nearly defect-free regions increases the degree of spin alignment in the defect-rich regions, thus increasing $T_C'$. Here, we hypothesize that $T_C'$ lies above the measurement temperature range in NCO films exhibiting entirely metallic $\rho-T$, whereas it lies below for those exhibiting entirely semiconducting $\rho-T$. Therefore, semiconducting $\rho-T$ typical of polycrystalline NCO can continuously change to non-monotonic $\rho-T$ toward metallic $\rho-T$ with current-induced $T_C'$ enhancement, and if such an evolution indeed occurs, it will provide a physical basis for establishing a spin-related transport model in polycrystalline NCO.

In this letter, we investigate the transport properties of a polycrystalline NCO film on an MAO buffer layer on a Si(001) substrate, and find that $\rho-T$ drastically changes from semiconducting to nearly metallic with increasing in-plane applied current $I$. Based on the characteristics, we propose a spin-related transport model that incorporates spatial, current, and temperature dependencies of the degree of spin alignment, and consistently explain $\rho-T$ results. Our model is further examined from resistivity−current ($\rho-I$) and current−voltage ($I-V$) characteristics at various temperatures and magnetoresistance (MR) measurements under an in-plane magnetic field $H_{\parallel}$.

A Si(001) substrate with a H-terminated surface was thermally cleaned at a substrate temperature $T_{sub}$ =750 °C for 10 min in an ultra-high-vacuum electron-beam evaporation chamber. Then, a 1-nm-thick MAO layer was deposited at $T_{sub}$ = 650 °C and characterized as amorphous from a halo-shaped reflection high-energy electron diffraction (RHEED) pattern. Subsequently, a 36-nm-thick NCO layer was deposited on MAO at $T_{sub}$ =315 °C under radio frequency (RF) oxygen plasma, while the oxygen pressure and RF power were maintained at $2\times10^{-3}$ Pa and 200 W, respectively. A concentric ring-shaped RHEED pattern confirmed polycrystalline NCO. After exposed to air, the sample was annealed for 30 min at 400 °C in a 1-atm oxygen atmosphere under $H_{\parallel}$ = 500 Oe. An out-of-plane $\theta$–$2\theta$ X-ray diffraction pattern identifies a single spinel phase (supplementary material, Fig. S1). The $M_s$ value of $1.5\mu_B/f.u.$ (supplementary material, Fig. S2), approximately 75% of the theoretical value, indicates a number of defects in the NCO film presumably near grain boundaries.

A Hall-bar device was fabricated by photo-lithography and ion milling. The resistance $R$ and $\rho$ were estimated using a four-terminal method, as shown in Fig. 1(a). Figure 1(b) shows $\rho-T$ curves from 4 to 290 K at $I$ = 100 nA–1 mA, corresponding to the current densities of 3.1 A/cm$^2$–31 kA/cm$^2$. At 100 nA, the curve shows entirely semiconducting $\rho-T$ and is well reproduced using both VRH and NNH models (supplementary material, Fig. S4). At 10 μA, the curve branches downward from the 100-nA curve below approximately 150 K and exhibits a peak at 15 K (gold arrow). At 100 μA, the curve shifts to lower $\rho$ over the entire temperature range with its peak at a higher temperature of 98 K (orange arrow). At 1 mA, the curve shifts to even lower $\rho$ and has its peak at an even higher temperature of 164 K (red arrow), approaching entirely metallic $\rho-T$. The enlarged linear plot is shown in Fig. 1(c). The $\rho$ value at 290 K reaches 0.7 mΩ cm, which is noticeably low compared with previously reported room-temperature $\rho$ values of 0.8–2.3 mΩ cm[15,20,30].

Here, we propose a spin-related transport model that explains the drastic current-induced evolution from entirely semiconducting, to non-monotonic, and toward entirely metallic $\rho-T$ in Fig. 1 (b), in a unified way.

Figure 2(a) illustrates typical local spin directions (blue arrows) in adjacent polycrystalline NCO grains, where the shading intensity represents the randomness of spin alignment. The interfacial region near the grain boundary contains more defects than the grain bodies and therefore has a lower degree of spin alignment. Figure 2(b) illustrates the corresponding typical electron transport based on two considerations: (1) the energy barrier for electron transfer between Ni cations, shown in magenta, is governed by the local degree of spin alignment: a lower degree (more random) of spin alignment reduces the transfer integral and raises the barrier[27]. The estimated barrier height under the condition giving the lowest degree of spin alignment in our measurements (100 nA, 250–290 K), as discussed later, was 46.6 meV (supplementary material, Sec. III); and (2) the electron transport mechanism depends on both barrier heights and thermal assist: electrons undergo metallic conduction (orange arrow) when the barrier is almost flat, whereas they hop over the barrier with thermal assist (green arrow) or tunnel through it when thermal assist is insufficient (blue arrow).

From these considerations, spatial, current, and temperature dependences of the degree of spin alignment and electron transport were modeled. Figure 3 illustrates the spin and electron-transport configurations, where the sequences (i)→(ii-a)→(iii-a)→(iv) and (iii-a)→(iii-b)→(iii-c) respectively exemplify the effects of increasing $T$ and $I$. The inset in Fig. 3 shows the configurations mapped onto a temperature–spin-current plane.

*Spatial dependence*: The spatial inhomogeneity of the degree of spin alignment gives $T_{\mathrm{C,int}} \leq T_{\mathrm{C,b}}$, where $T_{\mathrm{C,int}}$ and $T_{\mathrm{C,b}}$ denote the Curie temperatures of the interfacial regions and grain bodies, respectively. Note that $T_{\mathrm{C,int}}$ corresponds to $T_{\mathrm{C}}'$. In the absence of $I$, $T_{\mathrm{C,b}}$ was estimated to be 250 K (supplementary material, Fig. S3). For $T \ll T_{\mathrm{C,b}}$, each grain body does not have a clear magnetic domain wall inside itself, since the grain diameter was estimated to be ~20 nm from atomic force microscopy measurements. Spins are therefore well-aligned in each grain body, yielding a finite average spin orientation $\langle \mathbf{s} \rangle$ of each grain, whereas $\langle \mathbf{s} \rangle = 0$ for $T_{\mathrm{C,b}} < T$. Likewise, in each interfacial region, spins are well-aligned for $T \ll T_{\mathrm{C,int}}$ and misaligned for $T_{\mathrm{C,int}} < T$.

*Spin-current dependence*: A spin-polarized current is injected from a grain body into the adjacent interfacial region. A larger spin current enhances the degree of spin alignment in the interfacial region and thereby reduces barrier heights there, leading to an increase in $T_{\mathrm{C,int}}$, as shown in the inset of Fig. 3. The reduction in barrier heights can also be interpreted as an effective reduction in the width $w_{\mathrm{int}}$ of each interfacial region, corresponding to the transition from (iii-a) toward (iii-c) through (iii-b).

*Temperature dependence*: There are two competing effects. First, increasing $T$ reduces the degree of spin alignment, increasing barrier heights. Starting from (i) at $T \ll T_{\mathrm{C,int}}$, increasing $T$ first reduces the degree of spin alignment in the interfacial regions, causing a transition to (ii-a) for $T <$ $T_{\mathrm{C,int}}$ and to (iii-a) for $T_{\mathrm{C,int}} < T$, and subsequently reduces that in the grain bodies, causing a transition to (iv) for $T_{\mathrm{C,b}} < T$. Second, increasing $T$ allows electrons to hop over higher barriers. The competition between these two effects leads to a $\rho-T$ curve with its peak near $T_{\mathrm{C,int}}$, as reported in LSMO[27]. In addition, increasing $T$ reduces the magnitude of the spin current into the interfacial regions through the reduction in $\langle \mathbf{s} \rangle$ in each grain body, the spin source.

The current-induced evolution of $\rho-T$ in Fig. 1(b) is explained by the configurations in Fig. 3 and their map in the inset, where $T_{\mathrm{C,int}}$ and $T_{\mathrm{C,b}}$ increase with increasing spin current. At 100 nA, the trajectory with increasing $T$ proceeds from (iii-a) to (iv), as shown in the first row of the inset map, with $T_{\mathrm{C,int}}$ below the measurement temperature range of 4–290 K. The transport mechanism is governed by VRH at low temperatures in (iii-a) and NNH in (iv) (supplementary material, Sec III). At 10 μA, the trajectory with increasing $T$ proceeds from (ii-b) to (iii-b) with $T_{\mathrm{C,int}}$ emerging ∼15 K, merges with the 100-nA trajectory at 150 K and then proceeds through (iii-a) to (iv). With increasing $I$ from 100 nA to 10 μA, the current-induced change vanishes above 150 K because each grain body has reduced $\langle \mathbf{s} \rangle$ and generates a relatively smaller spin current than below 150 K. At 100 μA and 1 mA, the trajectories with increasing $T$ proceed from (ii-b) to (iii-b) and approach (iv), while they keep away from the 100-nA trajectory in the entire measurement temperature range, with $T_{\mathrm{C,b}}$ increased above 290 K. Regarding the interfacial regions, $T_{\mathrm{C,int}}$ increases from ∼15 K (10 μA) to ~98 K (100 μA) and ∼164 K (1 mA), while $w_{\mathrm{int}}$ decreases with increasing $I$. The $T_{\mathrm{C,int}}(=T_{\mathrm{C}}')$ enhancement plotted against current density is comparable to that reported for spin injection within an epitaxial LSMO nanowire[25] (supplementary material, Fig. S5), supporting our model based on spin-injection-induced spin alignment and indicating abundant spin-polarized conduction electrons in the grain bodies in our film.

Figure 4(a) shows $I-V$ characteristics from 4 to 290 K. Figure 4(b) shows the corresponding $\rho-I$ characteristics, each representing a constant-$T$ profile of $\rho-T$ characteristics in Fig. 1(b). At 10 K (purple), $\rho$ is nearly constant at $I <$ ∼1 μA and then decreases drastically with increasing $I$, indicating a change in the transport mechanism. The critical current $I_{\mathrm{c}}$ is defined here as the onset $I$ value of this decrease. As $T$ is increased, $\rho$ below $I_{\mathrm{c}}$ decreases, and $I_{\mathrm{c}}$ shifts higher. When the constant $\rho$ values in the flat part below $I_{\mathrm{c}}$ were extrapolated to 100 nA where inaccurate or missing points are present, and plotted as a function of temperature, the $\rho-T$ curve at 100 nA in Fig. 1(b) is well reproduced (supplementary material, Fig. S6(a)). Thus, the flat part corresponds to VRH or NNH in (iii-a) and (iv). Above $I_c$, the drastic decrease in $\rho$ is attributable to the decrease in $w_{int}$ since the configurations for 10 μA–1 mA at 10 K are (ii-b), where $w_{int}$ progressively decreases with increasing $I$.

The current-induced change in electron transport within (ii-b) above $I_c$ is further analyzed using the $I-V$ characteristics in Fig. 4(a). They all exhibit a multi-step structure, while the position of each step varies with temperature. The enlarged $I-V$ curve at 10 K in Fig. 4(c) is divided into regimes (A)−(D) according to its slope; each regime reflects the evolution of configurations as follows. At 10 K, the part around $I_c \sim 1$ μA corresponds to (B), whereas that above $I_c$ within (ii-b) corresponds to (C) and (D). The rapid increase in $I$ with increasing $V$ in (C) is attributed to a significant reduction in $w_{int}$ associated with the evolution from (iii-a) to (ii-b). The gradual increase in $I$ with increasing $V$ in (D) is attributed to an approach toward (iii-c), where metallic electron transport becomes dominant and the decrease in $\rho$ with increasing $I$ begins to saturate. Regarding the temperature dependence, Fig. 4(a) shows that the boundary $I$ between (B) and (C) and that between (C) and (D) both increase with increasing $T$. This feature confirms that the efficiency of current-induced enhancement of the degree of spin alignment decreases at higher $T$, owing to decreased $\langle \mathbf{s} \rangle$ of each grain body.

Figure 5(a) upper half shows $\Delta R = R - R_{\mathrm{peak}}$ plotted against $H_{\parallel}$ (MR curve) at 100 nA in (A) measured at 10 K, where $R_{\mathrm{peak}}$ is defined as the average $R$ at the two peaks. The condition of 10 K and $I \leq 100$ nA corresponds to (iii-a). The lower half shows $M-H_{\parallel}$ without $I$, where red and blue curves respectively represent the positive and negative sweeps between ±3 kOe. The $M-H_{\parallel}$ curve exhibits a clear hysteresis with its coercive force $H_c$ at approximately 0.5 kOe (dashed lines). Considering that $M-H_{\parallel}$ primarily reflects the degree of spin alignment in the grain bodies that occupy most of the film, $\langle \mathbf{s} \rangle$ is random across grains when $M = 0$ at $H = H_c$, whereas with increasing $H_{\parallel}$ from $H_c$, $\langle \mathbf{s} \rangle$ becomes progressively uniform across grains. Regarding the interfacial regions, the spin alignment in each region is presumably random near $H_c$ and becomes more aligned with increasing $H_{\parallel}$, accordingly. The MR curve shows that $\Delta R$ is highest at $H_c$ and decreases with increasing $H_{\parallel}$. The negative MR can be simply explained by the transition from (iii-a) near $H_c$ toward (iii-b) with increasing $H_{\parallel}$, where the magnetic-field-induced spin alignment in the interfacial regions reduces hopping barriers, leading to a decrease in $R$. Such negative MR was commonly seen in previous reports[15,31]. Figure 5(b) shows $\Delta R_{3\mathrm{k}} = R_{\pm 3\mathrm{k}} - R_{\mathrm{peak}}$ at 10 K plotted against $I$ from 100 nA to 100 μA, where $R_{\pm 3\mathrm{k}}$ is defined as average $R$ at ±3 kOe. The magnitude of the negative MR decreases from 100 nA to 1 μA. At 1 μA in (B), owing to a certain degree of spin-alignment even without $H_{\parallel}$, the additional contribution from magnetic-field-induced spin alignment is small, similar to the origin of decreased magnitude of negative MR in higher-quality NCO films[21].

A distinct sign reversal of MR emerges when $I$ is further increased to 10 μA in (C). At 100 μA in (C), the sign remains positive with smaller $\Delta R_{3\mathrm{k}}$. This sign reversal between (B) and (C) is very likely associated with some microscopic change in the interfacial regions (supplementary material, Secs. VIA-C) and supports the transition from (ii-a) toward (iii-c) as $I$ crosses the boundary between (B) and (C). Although the mechanism for the positive MR is unclear at this stage, one possible scenario involves the generation of up-spin electrons in the interfacial regions (supplementary material, Sec. VID).

In conclusion, the systematic investigation of the current-dependent $\rho{-}T$ in the polycrystalline NCO film has clarified the spin-related transport properties over wide temperature and current ranges, based on our transport model that extends the conventional hopping models via incorporation of spatial defect and spin-alignment distributions. Abundant spin-polarized conduction electrons are present in the grain bodies, highlighting the potential for an efficient spin-injection source for Si. With increasing in-plane current, more spin current is injected from the grain bodies to the near-grain-boundary regions, where the initially low degree of spin alignment increases and the initially high hopping barriers are reduced through strengthened double-exchange interactions. This allows the metallic transport of the grain bodies to emerge in $\rho{-}T$. With the utilization of the coexistence of contrasting magnetic orders inherent to polycrystalline materials, this study advances the understanding of the spin-related transport properties of NCO, thereby contributing to a broader understanding of the physics of double-exchange systems.

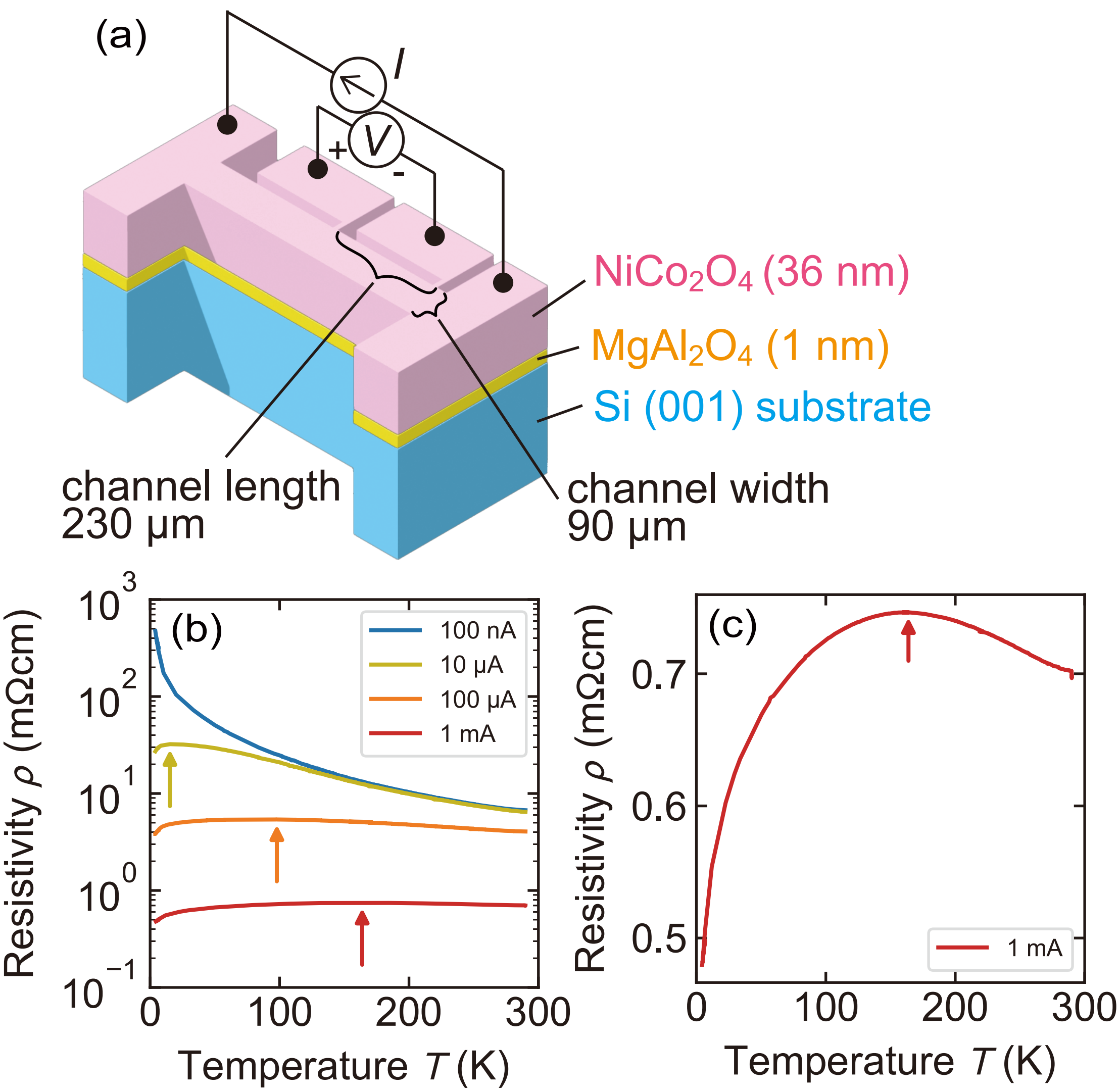

FIG. 1. (a) Schematic illustration of a Hall-bar device and the vertical sample structure. (b) Resistivity *vs*. temperature ($\rho-T$) characteristics measured at a constant current $I$ (= 100 nA−1 mA), where an arrow indicates the peak of each curve. (c) Linear plot of $\rho-T$ at $I$ = 1 mA.

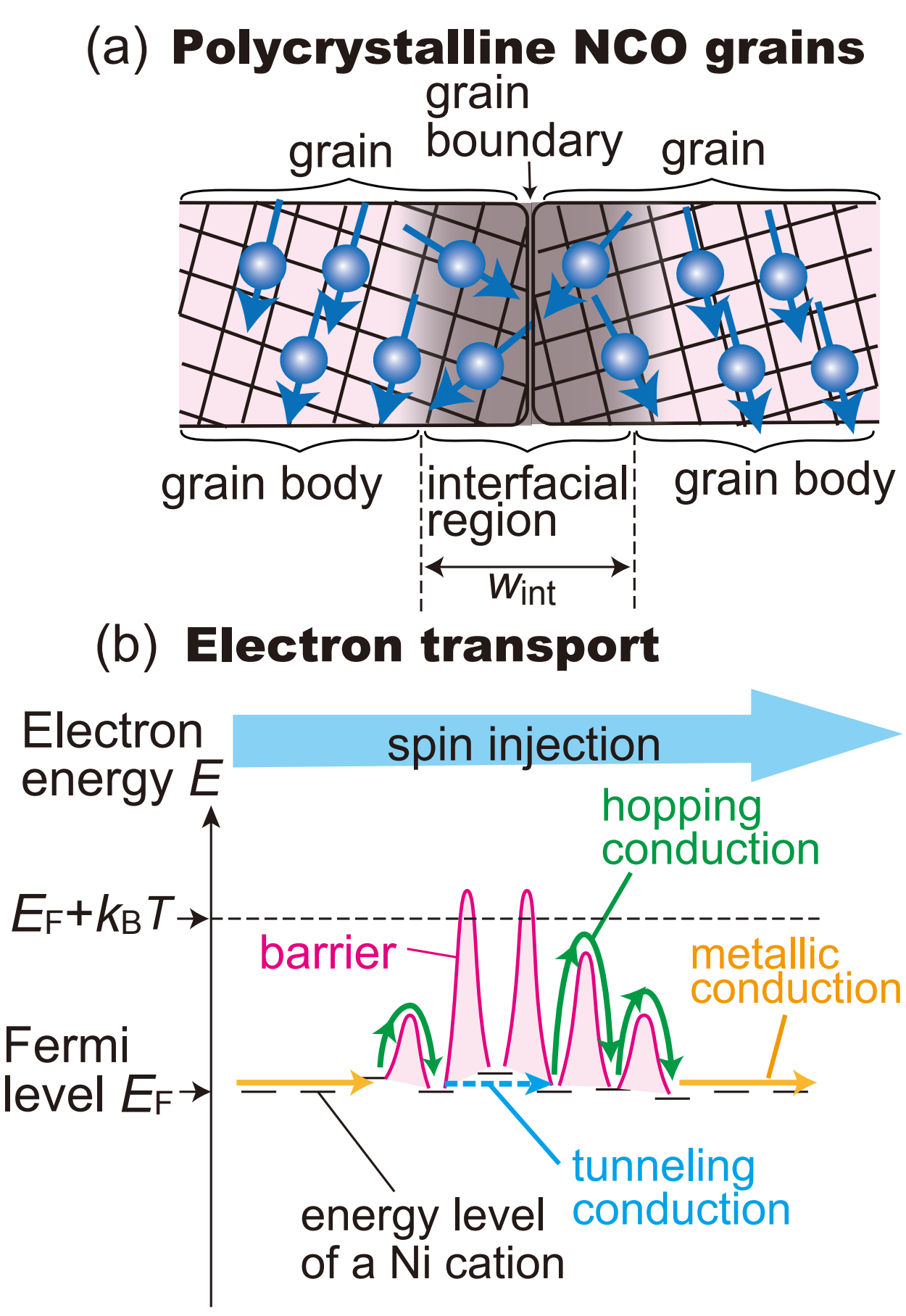

FIG. 2. Schematic illustrations of (a) polycrystalline NCO grains and local spin directions (blue arrows) and (b) electron transport including hopping, tunneling, and metallic conduction.

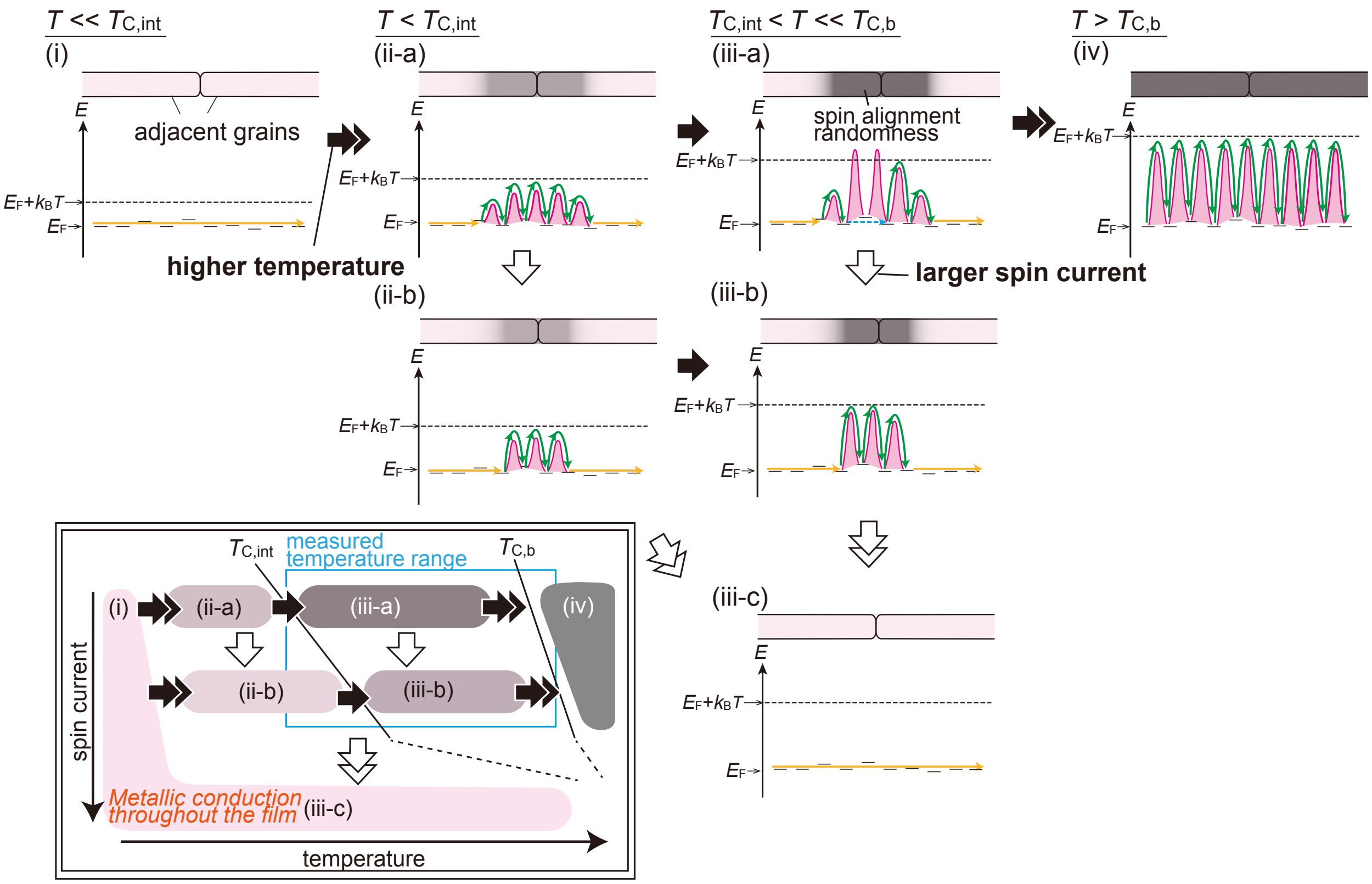

FIG. 3. Spin-related transport configurations under varying temperature and spin-current. The inset shows the configurations mapped onto a temperature–spin-current plane.

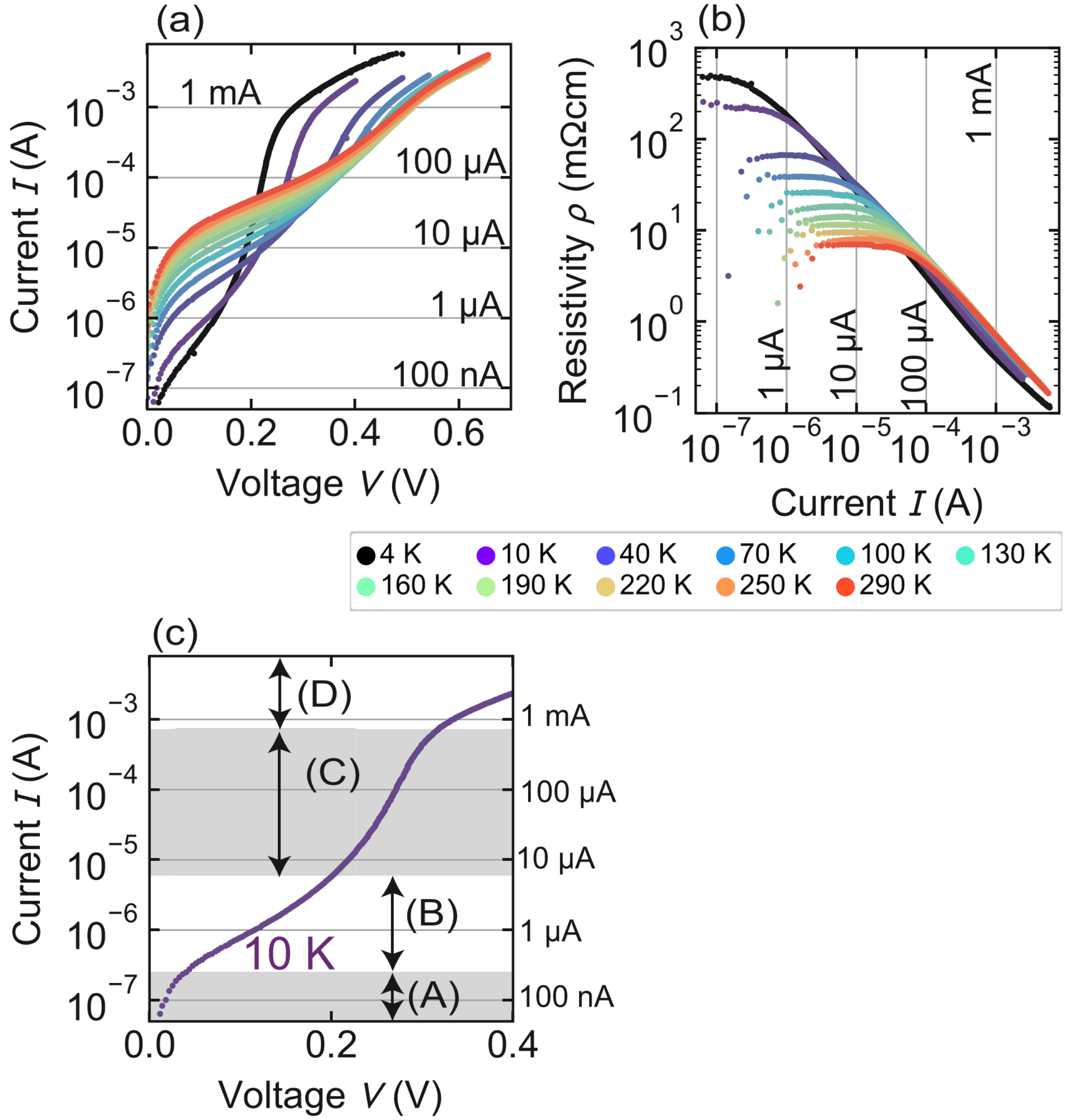

FIG. 4. (a) Current-voltage ($I-V$) characteristics measured at a constant temperature $T$ (=4−290 K). (b) Resistivity−current ($\rho$–$I$) characteristics obtained from (a). (c) Enlarged view of the $I-V$ curve at 10 K, which is divided into regimes (A)−(D) according to its slope.

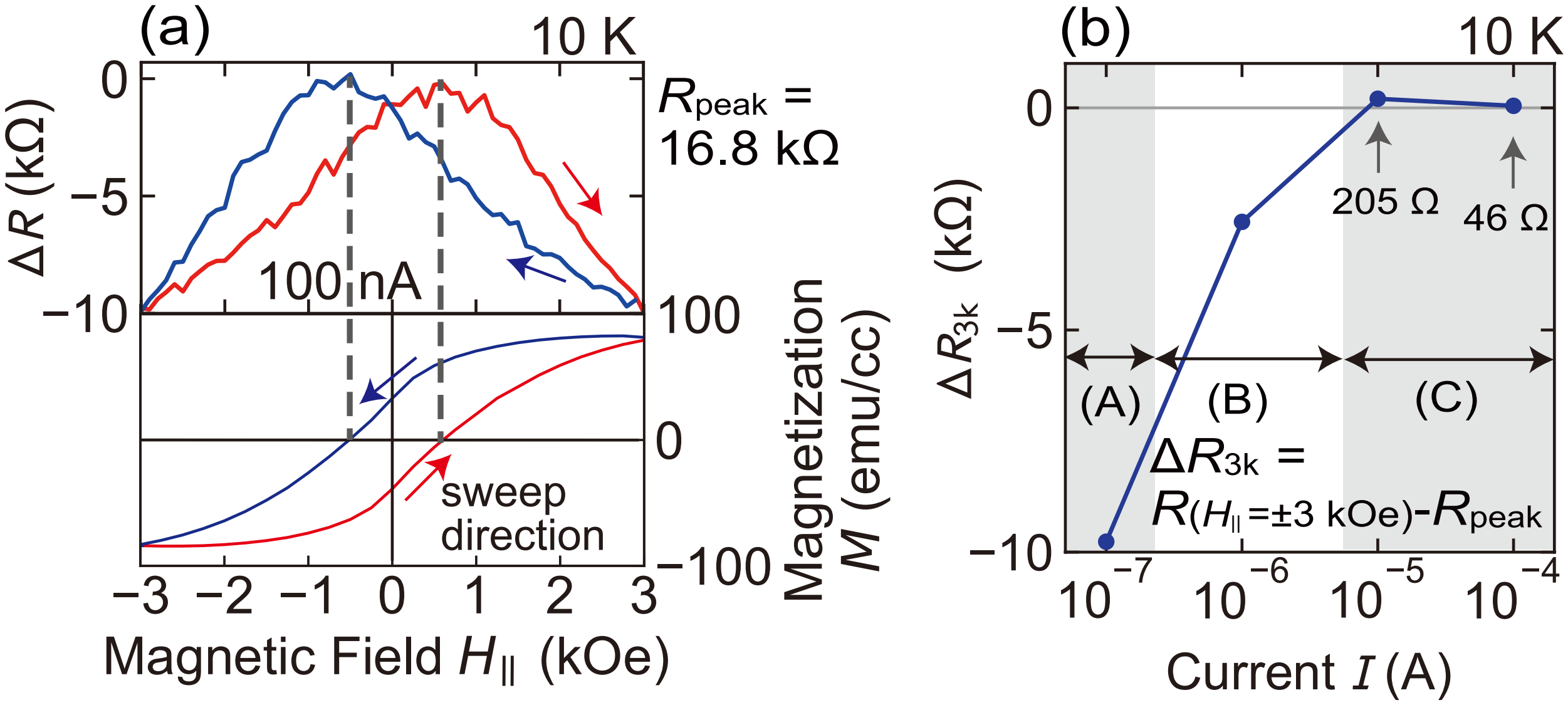

FIG. 5. (a) Magnetoresistance curve with $I$ = 100 nA and magnetization curve without $I$, both at 10 K. (b) $\Delta R_{3k}$ at 10 K plotted against $I$.

## SUPPLEMENTARY MATERIAL

See the supplementary material for additional experimental details.

## ACKNOWLEDGMENTS

The authors express their gratitude to Dr. Baisen Yu (The Univ. of Tokyo) for his valuable contribution during the early stages of this research and Dr. Shoichi Sato for helpful discussion. This work was partly supported by Grants-in-Aid for Scientific Research (20H05650, 23K17324, 23H00177, and 25H00840) and Spintronics Research Network of Japan (Spin-RNJ).

## AUTHOR DECLARATIONS

### Conflict of Interest

The authors have no conflicts to disclose.

### Author Contributions

**Shiho Sugiyama:** Conceptualization (equal); Data curation (equal); Investigation (equal); Methodology (equal); Writing – original draft. **Masaaki Tanaka:** Conceptualization (equal); Investigation (equal); Methodology (equal); Writing – review & editing (equal); Funding acquisition (equal). **Ryosho Nakane:** Conceptualization (equal); Data curation (equal); Investigation (equal); Methodology (equal); Writing – review & editing (lead); Funding acquisition (equal); Supervision (lead).

## DATA AVAILABILITY

The data that support the findings of this study are available from the corresponding author upon reasonable request.

# SUPPLEMENTARY MATERIAL

**Spin-related transport in a polycrystalline $NiCo_2O_4$ film: Drastic current-induced change in resistivity–temperature characteristics via spin injection**

Shiho Sugiyama[1], Masaaki Tanaka[1,2,3], and Ryosho Nakane[1,3,4]

[1]Department of Electrical Engineering and Information Systems, The University of Tokyo, 7-3-1 Hongo, Bunkyo-ku, Tokyo 113-8656, Japan

[2]Center for Spintronics Research Network (CSRN), The University of Tokyo, 7-3-1 Hongo, Bunkyo-ku,Tokyo 113-8656, Japan

[3]Institute for Nano Quantum Information Electronics, The University of Tokyo, 4-6-1 Komaba, Meguro-ku, Tokyo 153-8505, Japan

[4]Systems Design Lab (d.lab), The University of Tokyo, 7-3-1 Hongo, Bunkyo-ku, Tokyo 113-8656, Japan

## I. Structural characterization

The crystallographic phase of our $NiCo_2O_4$ (NCO) film was identified from the peak positions in an X-ray diffraction pattern detected by an out-of-plane $\theta$–$2\theta$ scan using a Cu Kα1 X-ray ($\lambda$ = 0.15406 nm) source. The main purpose here is to examine whether decomposition occurred in our film; NCO can decompose into antiferromagnetic insulators NiO and/or CoO under certain conditions such as oxygen deficiency or high temperatures[S1]. Figure S1 shows the XRD pattern. All observed diffraction peaks were assigned to NCO in a spinel-type structure[S2], except for the peak indicated by a black triangle, which is the satellite of the NCO (311) peak caused by the diffraction with a Cu Kβ X-ray ($\lambda$ = 0.13922 nm). The absence of diffraction peaks corresponding to NiO or CoO[S3,S4] indicates no detectable decomposition in the film.

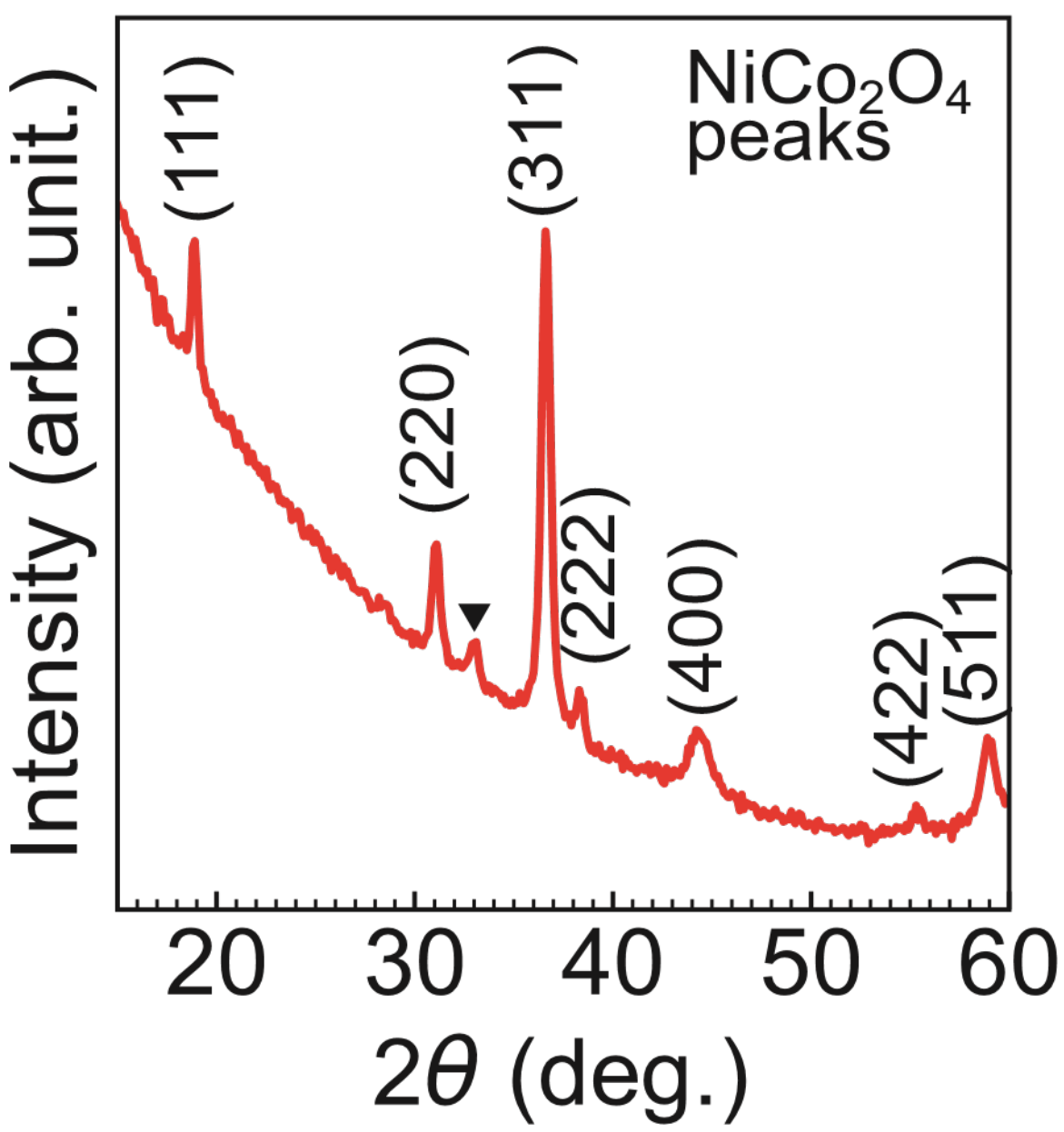

FIG. S1. Out-of-plane $\theta$-$2\theta$ scan, where peak assignments are indicated by black labels and a satellite peak of the NCO (311) peak from the Cu Kβ X-ray is indicated by a black triangle.

## II. Magnetic characterization

Magnetic properties were characterized using a superconductive quantum interference device magnetometer under an in-plane magnetic field $H_{\parallel}$. The diamagnetic susceptibility of the Si substrate was estimated from the negative linear slope of the measured magnetic moment against $H_{\parallel}$, assuming that the magnetization $M$ of the NCO film is saturated at higher $H_{\parallel}$. Figure S2 shows the magnetization *vs*. in-plane magnetic field ($M$−$H_{\parallel}$) curve of our NCO film. The saturation magnetization was estimated to be $1.5\mu_{\mathrm{B}}/f.u.$ Figure S3 shows a magnetization−temperature ($M$−$T$) curve under $H_{\parallel}$ = 250 Oe, where the measurement during heating was performed after cooling under $H_{\parallel}$ = 20 kOe. The Curie temperature $T_{\mathrm{C,b}}$ of the grain bodies, which constitute the dominant part of the NCO film, was estimated to be approximately 250 K, at which $M$ levels off.

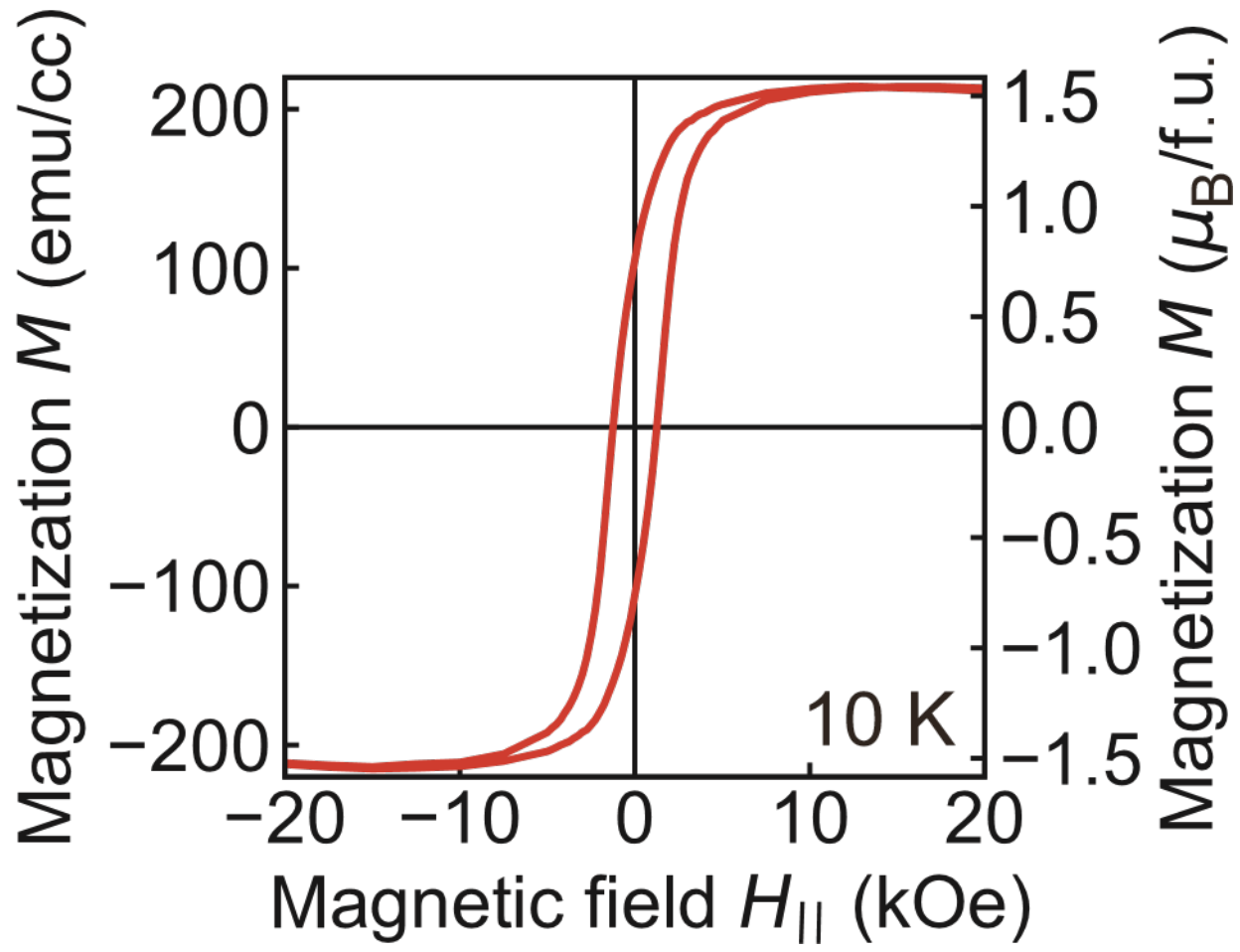

FIG. S2. Magnetization *vs*. in-plane magnetic field ($M$−$H_{\parallel}$) curve at 10 K, where the magnetization $M$ is shown in units of emu/cc on the left vertical axis and $\mu_B$/*f.u.* on the right vertical axis.

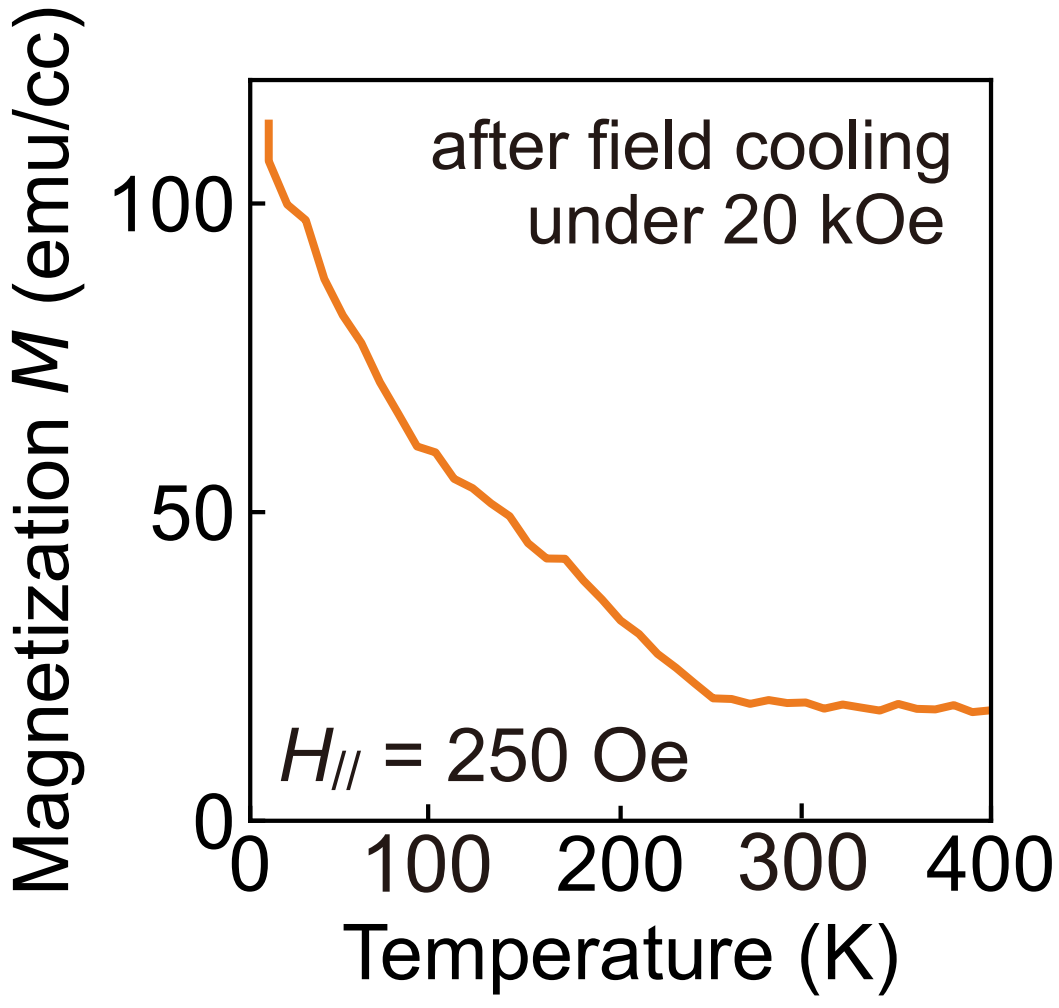

FIG. S3. Magnetization $M$ under $H_{\parallel}$ = 250 Oe plotted as a function of temperature, where the measurement during heating was performed after cooling under $H_{\parallel}$ = 20 kOe.

## III. Fitting analysis of the temperature-dependent transport at 100 nA

The transport mechanism of our NCO film at $I$ = 100 nA in Fig. 1(b) in the main text was analyzed by the fitting of the Mott-3D variable range hopping (VRH) and nearest neighbor hopping (NNH) conduction models[S5] to the experimental temperature dependence of conductivity ($\sigma$−$T$). Assuming that the carrier density is independent of temperature, the conductivity is described as

$$\sigma = \sigma_0 exp\left[-\left(\frac{T_{\mathrm{VRH}}}{T}\right)^{\frac{1}{4}}\right], \sigma_0 \propto \frac{1}{T^{\frac{3}{2}}}$$

when the VRH is dominant, and when the NNH is dominant as

$$\sigma = \sigma_0 \exp\left(-\frac{T_{\mathrm{NNH}}}{T}\right), \sigma_0 \propto \frac{1}{T}$$

where $T_{\mathrm{VRH}}$ and $T_{\mathrm{NNH}}$ denote the characteristic temperatures of the VRH and NNH models, respectively. Figure S4 shows the $\sigma-T$ curve at 100 nA corresponding to $\rho-T$ in Fig. 1(b) in the main text, together with the fitting curves. The experimental data were fitted by the VRH model in the range of $T$ = 35‑100 K (green shaded region), yielding $T_{\mathrm{VRH}} = 6.72 \times 10^5$ K, and by the NNH model in the range of $T$ = 250‑290 K (orange shaded region), yielding $T_{\mathrm{NNH}}$ = 541 K (46.6 meV). The experimental data are well reproduced by these models. The intersection temperature of the two fitting curves is 149 K. Note that the characteristic temperatures reflect the highest temperatures along the transport paths among the heights presumably varying within the interfacial regions, while the main focus in our paper is inter-region variations rather than such intra-region variations.

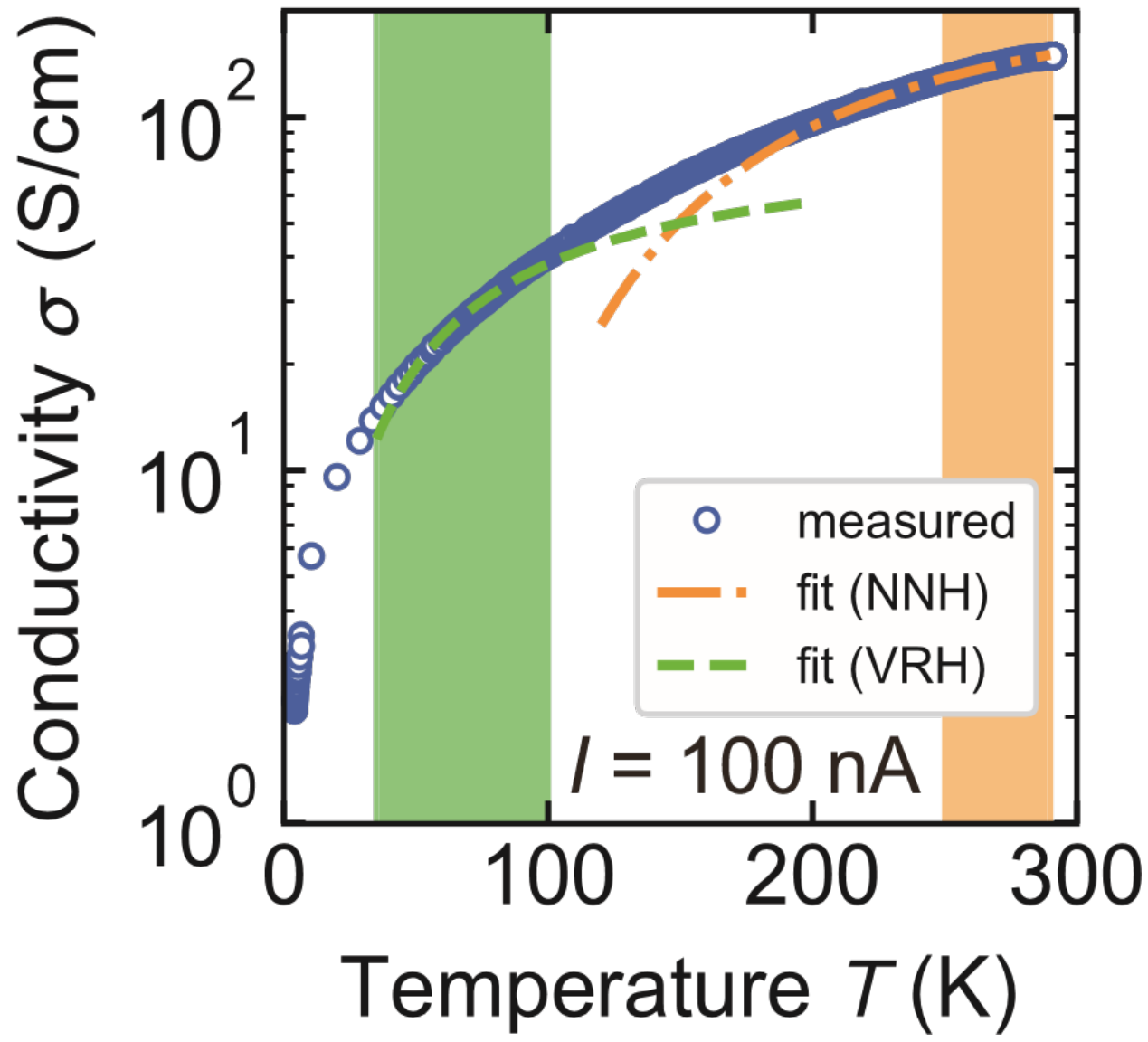

FIG. S4. Temperature dependence of conductivity ($\sigma-T$) at $I$ = 100 nA, where the blue open circles denote the experimental data and the green dashed and orange dash-dotted curves denote the fitting curves of the VRH and NNH models, respectively. The green and orange shaded regions represent the temperature ranges used for the VRH and NNH fittings, respectively.

## IV. Discussion on the high-field VRH conduction

In Figure 1(b) in the main text, the temperature dependence of resistivity ($\rho - T$) continuously changes with increasing in-plane current from 100 nA to 1 mA: the steep, exponential increase in resistivity with decreasing temperature observed at 100 nA is progressively suppressed, resulting in the much weaker temperature dependence at 1 mA. Here, high-field VRH conduction[S6] is excluded from the origin of this change, where the electric field dominates conduction and resistivity becomes independent of temperature. This is because the electric field at 1 mA is only 0.02 kV/cm, well below the field range for the development of this conduction.

## V. Comparison with a previous study on the current-induced Curie-temperature enhancement in $La_{0.7}Sr_{0.3}MnO_3$

Our study has reported current-induced enhancement of the Curie temperature $T_{C,int}$ of the interfacial regions near grain boundaries in the current-density range of 0.31–3.1 kA/cm$^2$, which is attributed to spin injection that enhances the degree of spin alignment there. To examine the validity of this model in this current-density range, our results are compared with the related study by Ono *et al.*[S7] In their study, a 50-nm-thick and 10-μm-wide epitaxial $La_{0.7}Sr_{0.3}MnO_3$ (LSMO) wire was partially irradiated with $Ga^+$ ions to locally decrease the Curie temperature through the introduction of many defects, and prepared a sequential ferromagnet/paramagnet/ferromagnet structure along the longitudinal direction of the wire. They reported that with increasing current through the wire, the locally decreased Curie temperature, $T_C'$, in the irradiated region increases owing to spin-injection from the adjacent non-irradiated ferromagnetic region. Figure S5 shows $T_{C,int}$ (orange squares in our work) and $T_C'$ (blue circles in Ref. S7) plotted against current density. In our work, $T_{C,int}$ was estimated from the temperature of the peak of the $\rho - T$ curves in Fig. 1(b) in the main text.

In the low-current-density regime, the rate of increase in $T_{C,int}$ with increasing current density is higher than that reported by Ref. S7. In the high-current-density regime, the rate of increase decreases and becomes comparable to that in Ref. S7. The current density range for the $T_{C,int}$ enhancement is comparable to that reported by Ref. S7, supporting the validity of our interpretation based on spin-injection-induced spin alignment. The comparable and/or higher rate of increase in $T_{C,int}$ in our polycrystalline NCO film relative to the epitaxial nanowire of LSMO, another theoretically half-metallic oxide, indicates that the grain bodies in our film are an efficient spin source with abundant spin-polarized conduction electrons, comparable to the non-irradiated and nearly defect-free epitaxial LSMO nanowire.

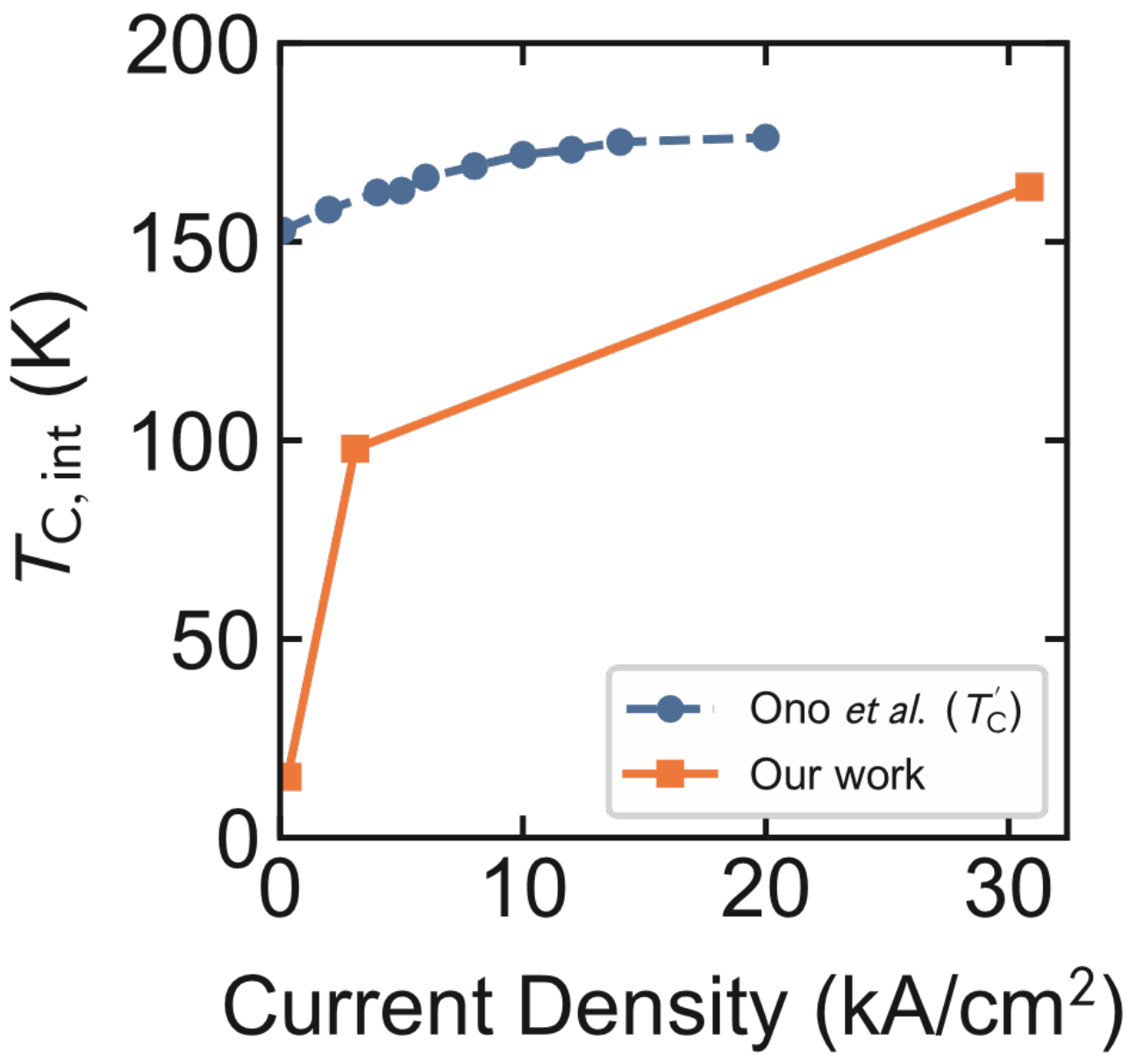

FIG. S5. Comparison of current-density enhancement of $T_C$, where orange squares correspond to the estimated $T_{C,int}$ of our NCO film whereas blue circles correspond to $T_C'$ of irradiated LSMO reported by Ono *et al.*[S7]

## VI. Consistency between the *ρ*–*I* extrapolation and *ρ*–*T*

The resistivity−current ($\rho$−$I$) characteristics in Fig. 4(b) in the main text have some inaccurate or missing points in a very low current range below $10^{-6}$ A, owing to the limitations of the experimental setup. To validate the extrapolation used for these regions, the extrapolated $\rho$−$I$ characteristics were compared with the corresponding $\rho$−$T$ in the main text, as shown in Figs. S6(a) and (b). Figure S6(a) shows $\rho$−$I$ characteristics in which $\rho$ is extrapolated toward lower $I$ as a constant value from the flat part for each $T$, as indicated by the horizontal dashed line. The extrapolated values along the $I$ =100 nA profile in Fig. S6(a) are shown by orange circles in Fig. S6(b), which agree very well with the $\rho$−$T$ curve in Figure 1(b) at $I$ = 100 nA (blue curve in Fig. 6(b)), validating our estimation method using the extrapolation.

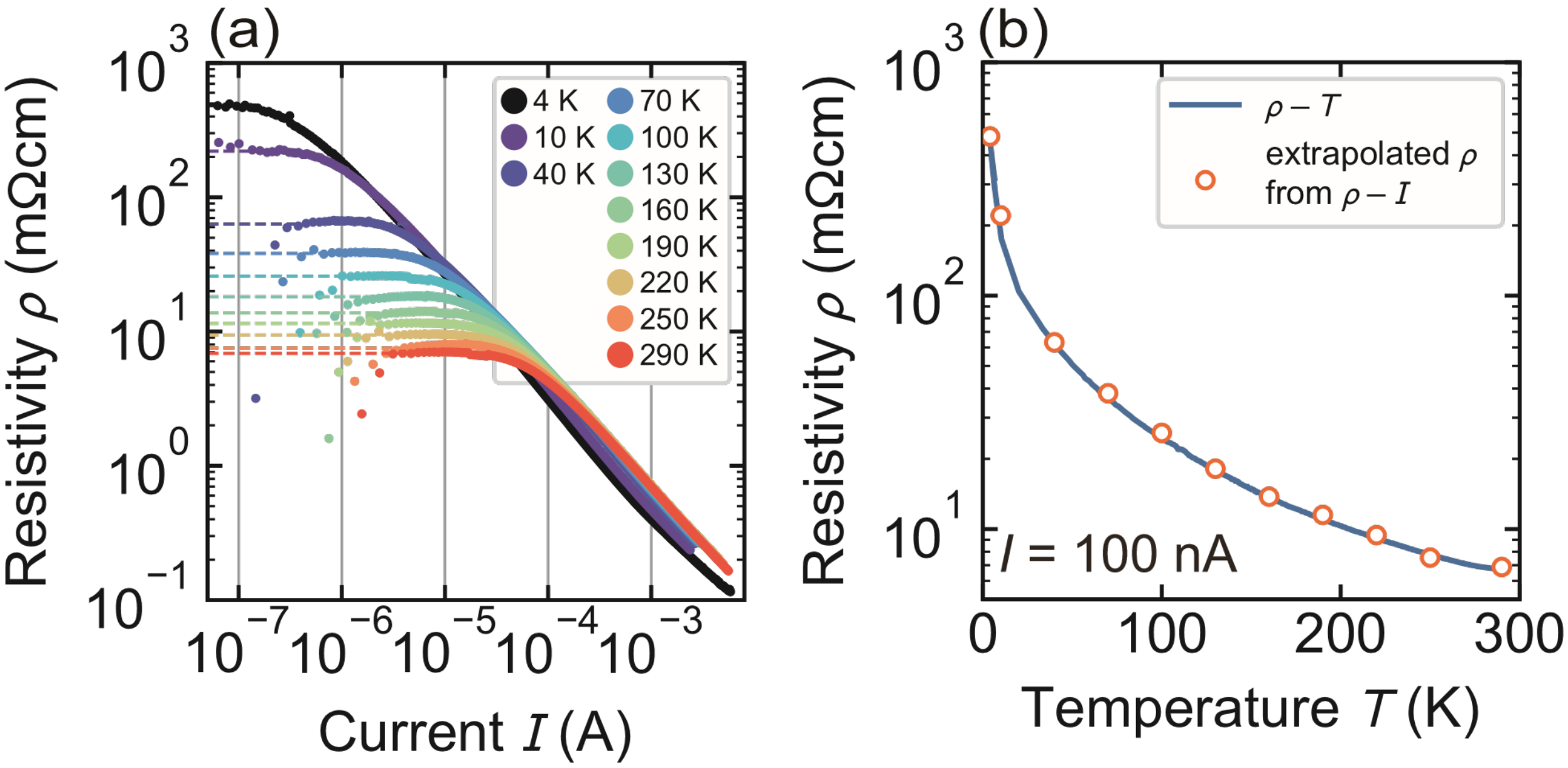

FIG. S6. (a) Resistivity-current ($\rho$−$I$) characteristics at constant $T$ ranging from 4–290 K, where each horizontal dashed line is the extrapolated $\rho$ value toward lower $I$ for each $T$ value. (b) Comparison of the extrapolated $\rho$ values at 100 nA from (a) (orange open circles) and the experimental $\rho$−$T$ data in Fig. 1(b) (blue curve).

## VII. Supplementary results and discussion on the current-induced sign reversal of magnetoresistance

Supplementary data and discussion on the current-induced sign reversal of magnetoresistance (MR) in Fig. 5 in the main text are provided. First, Joule heating is excluded as the origin of the sign reversal in Subsection A. Second, the reproducibility of the sign reversal is confirmed using $I$–$V$ characteristics in Subsection B. Third, the positive and negative MR curves are compared to reveal the similarities and differences in Subsection C. Finally, a possible microscopic mechanism for the positive MR is discussed in Subsection D.

### A. Exclusion of Joule heating as the origin of the sign reversal

Figure S7 shows $\Delta R_{3k}$ at 10 μA plotted against temperature ($T$ = 10–40 K), where $\Delta R_{3k}$ = $R$(±3 kOe) - $R_{peak}$. $\Delta R_{3k}$ is positive for these temperatures, while its magnitude decreases with increasing $T$. Therefore, the current-induced sign reversal of MR from negative to positive is not attributed to Joule heating caused by increasing current, because increasing $T$ decreases $\Delta R_{3k}$ rather than increasing it.

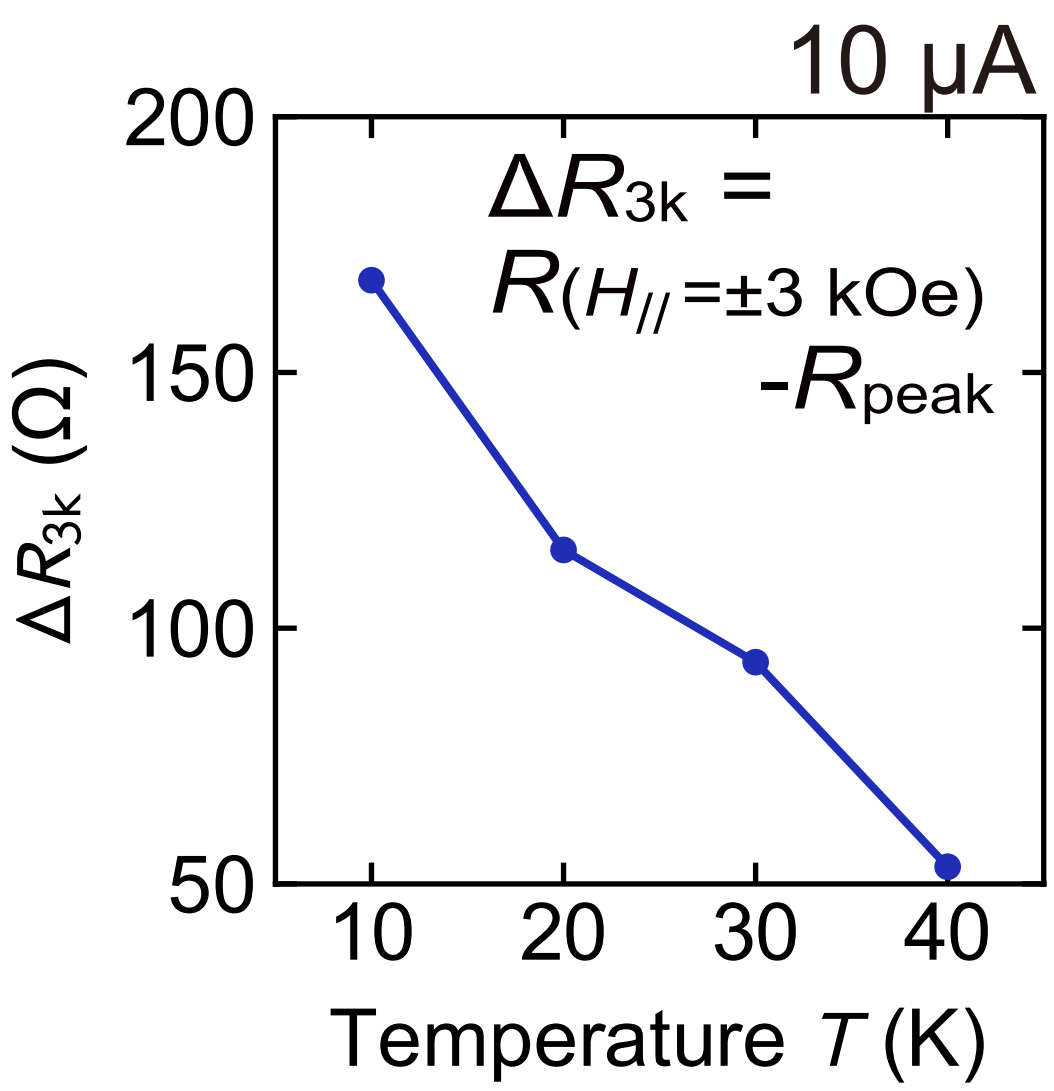

FIG. S7. $\Delta R_{3k}$ at 10 μA plotted against temperature $T$, where $\Delta R_{3k} = R(\pm 3\ \text{kOe}) - R_{peak}$.

## B. Reproducibility of the MR sign reversal in the $I$–$V$ characteristics

As discussed in the main text, the combined analysis of the current-voltage ($I-V$) characteristics in Fig. 4(c) and the current dependence of MR in Fig. 5(b) shows that the sign of MR is negative below and positive above the $I$ and/or $V$ values at which $I$ starts to drastically increase with increasing $V$, namely at the boundary between regimes (B) and (C). The $I-V$ curve is divided into four regimes (A)−(D) according to the slope as shown in Fig. 4(c) and Fig. S8(a).

This sign reversal of MR with increasing $I$ was also evaluated using $I-V$ characteristics under varying $H_{\parallel}$ (= 0, 1.5, 3 kOe). Figure S8(a) shows the $I-V$ characteristics for $V$ = 0–0.5 V. Figure S8(b) shows an enlarged view of (a) for $V$ = 0.05–0.25 V, where the boundary $V$ value between (B) and (C) is indicated by an orange triangle ($V$ = 0.18 V). Figure S8(c) shows $\Delta I \equiv I(H_{\parallel} = 0, 1.5, 3\ \text{kOe}) - I(H_{\parallel} = 0\ \text{Oe})$ plotted against $V$, which presents $I-V$ characteristics relative to those at $H_{\parallel} = 0$ Oe. The orange triangle indicates the same $V$ value as in (b). Note that a decrease in $\Delta I$, namely a decrease in $I$, for a fixed $V$ with increasing $H_{\parallel}$ corresponds to an increase in resistance, as indicated by the black arrow. At $V > 0.18$ V in (C), the resistance increases with increasing $H_{\parallel}$, exhibiting positive MR, whereas at $V < 0.18$ V in (B), the resistance decreases with increasing $H_{\parallel}$, exhibiting negative MR, except for some fluctuations. Hence, the sign reversal between the boundary between (B) and (C) is reproduced.

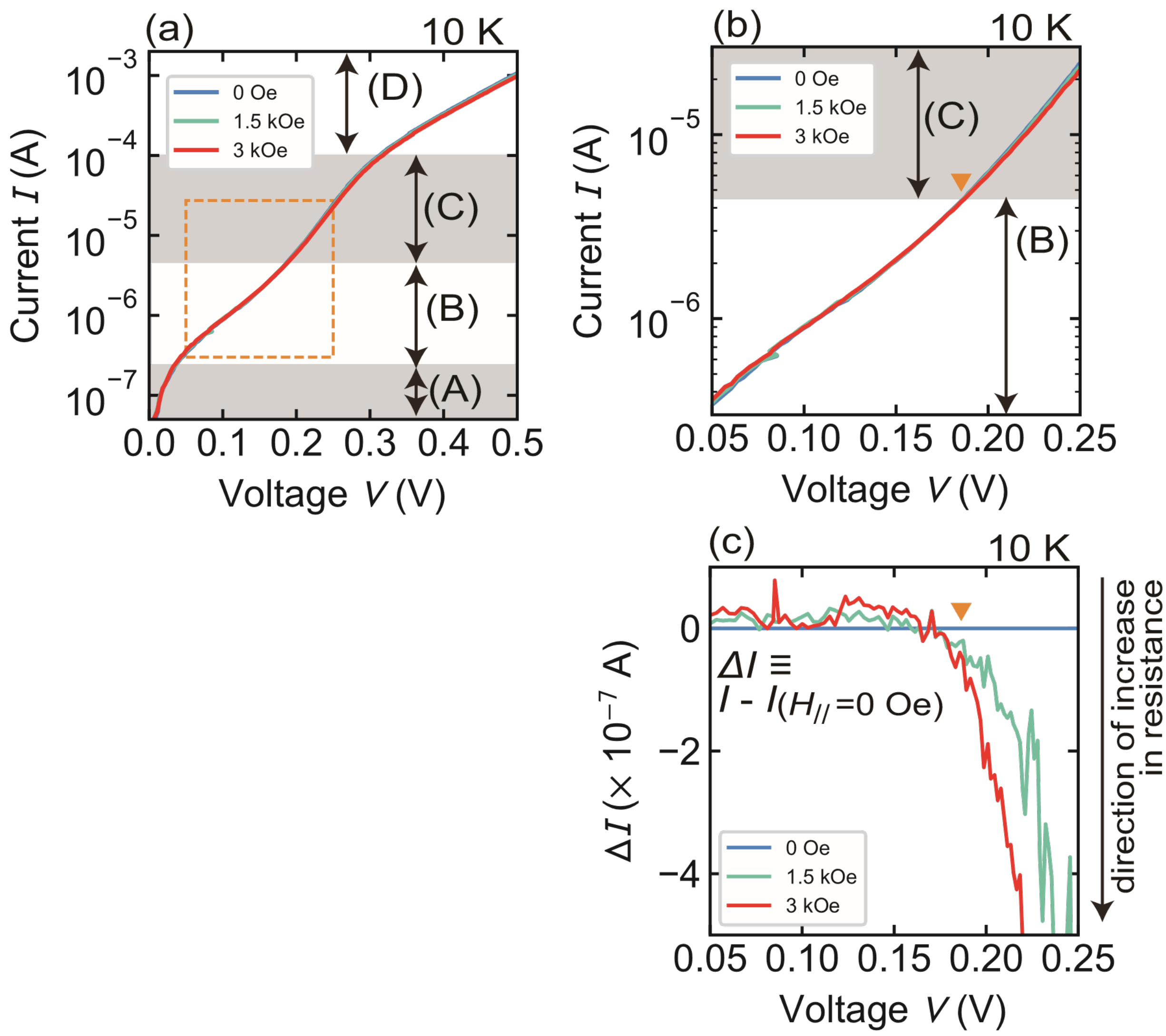

FIG. S8. (a) Current-voltage ($I$−$V$) characteristics at 10 K measured under a constant in-plane magnetic field $H_{\parallel}$ (= 0, 1.5, and 3 kOe). The curves are divided into regimes (A)−(D) according to their slopes. (b) Enlarged view of (a), where the $I-V$ range corresponds to the orange-dashed square in (a). The $V$ value of the transition between (B) and (C) is indicated by the orange triangle. (c) Current change $\Delta I \equiv I(H_{\parallel} = 0, 1.5,$ 3 kOe$) - I(H_{\parallel} = 0$ Oe$)$ plotted as a function of $V$, where the orange triangle is located at the same $V$ value as in (b) and the black arrow along the vertical axis denotes the direction corresponding to the decrease in $I$ at a fixed $V$ value, which is equivalent to the increase in resistance.

## C. Comparison between the negative and positive MR curves

The shapes of the positive MR curve at 1 mA and the negative MR curve at 100 nA are compared to investigate a possible difference in magnetic structure that may underlie the sign reversal of MR. Figure S9 shows MR at 1 mA in the upper half and MR at 100 nA in the lower half, with $H_{\parallel}$ swept between ±50 kOe. These curves exhibit almost the same saturation fields (dotted lines) and peak positions (dash-dotted lines), indicating that the spin alignment in grain bodies is nearly the same between these $I$ values. Therefore, some microscopic difference induced by $I$ is probably responsible for the sign reversal. Note that the difference in $R_{\text{peak}}$ at 100 nA between the measurements in Fig. S9 and Fig. 5(a) is attributed to the measurement history of the sample and does not affect the overall trend discussed in the main text.

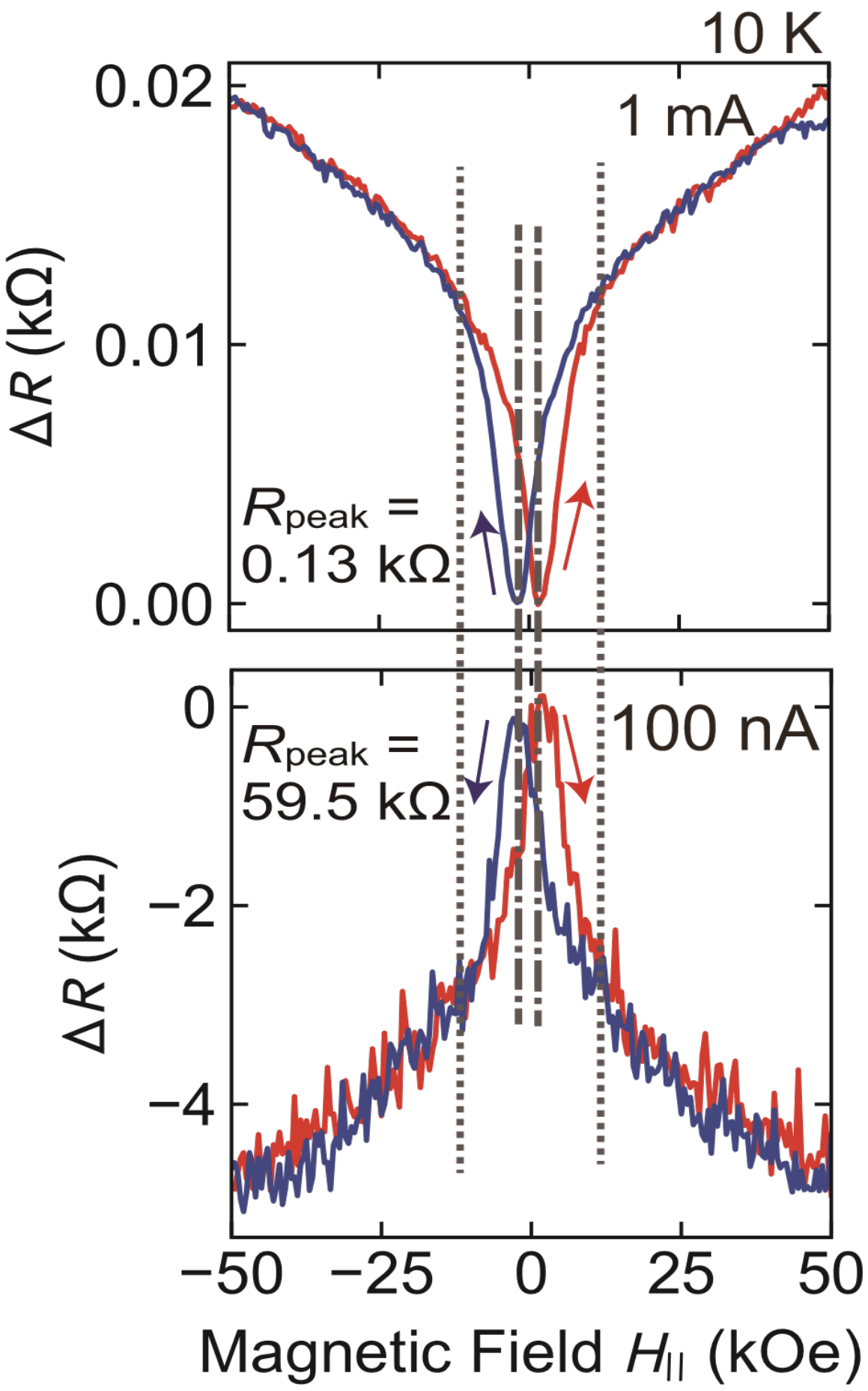

FIG. S9. MR curves at 10 K with $I = 1$ mA (upper half, positive MR) and 100 nA (lower half, negative MR).

## D. Possible origin of the positive MR

A possible scenario for the positive MR observed at higher currents is discussed here, although the mechanism remains unclear at this stage. In the ideal case, the density of states (DOS) of NCO has a band gap around the Fermi level $E_F$ for up-spin electrons, whereas it is continuous around $E_F$ for down-spin electrons, resulting in 100% spin-polarization of the transport carriers, as schematically illustrated in Fig. S10(a). Each grain body is very likely nearly half-metallic at 10 K, the temperature at which the positive MR was observed. In the interfacial regions near grain boundaries, oxygen vacancies introduce up-spin states in the band gap above $E_F$[S8], as indicated by

(1) in Fig. S10(b). The charging of defect states in the interfacial regions under a large current[S9] shifts the local $E_F$ upward so that it lies in the up-spin conduction band, inducing up-spin transport electrons there, as indicated by (2) in Fig. S10(b). This charging effect may also account for the multi-step $I$−$V$ characteristics. When $H_{\parallel}$ is sufficiently large, the magnetic moments of both the interfacial regions and grain bodies are aligned along the direction of $H_{\parallel}$. When an up-spin transport electron induced in an interfacial region enters the adjacent grain body, spin-flip scattering is unavoidable, since only down-spin states are available in the grain body around $E_F$ in this case. When $H_{\parallel}$ is reduced and the magnetic moments become less uniformly aligned, the states in the grain bodies acquire a finite up-spin component around $E_F$. As a result, up-spin electrons in an interfacial region can enter the adjacent grain body with a lower probability of spin-flip scattering. When $H_{\parallel}$ is decreased to the coercive force $H_c$, the magnetic moments of the individual regions become most randomly oriented with respect to each other, leading to the minimum probability of spin-flip scattering.

Hence, at higher currents, the spin-flip scattering rate is minimum at $H_{\parallel} = H_c$, giving rise to resistance minima around $H_c$, and increases with increasing $H_{\parallel}$, resulting in positive MR, as shown in the MR curve at 1 mA in the upper half of Fig. S9.

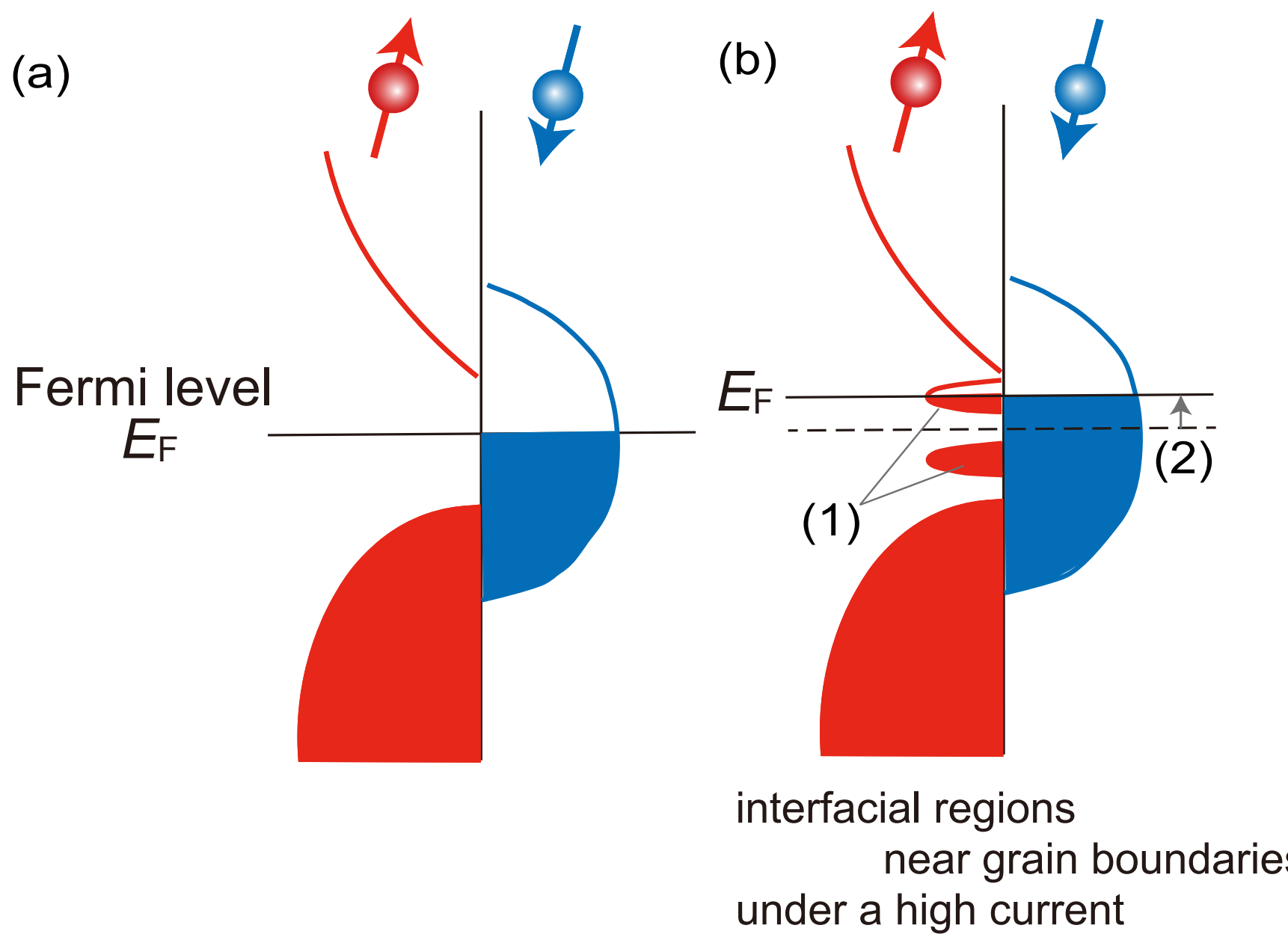

FIG. S10. (a) Schematic illustration of the spin-resolved density of states (DOS) in half-metallic NCO. Each grain body is very likely nearly half-metallic at 10 K, the temperature at which the positive MR was observed. (b) Schematic illustration of the spin-resolved DOS in the interfacial regions near grain boundaries under a high current in our scenario, where (1) up-spin states are introduced in the band gap and (2) the local Fermi level $E_F$ is shifted upward.

## REFERENCES FOR SUPPLEMENTARY MATERIAL